\documentclass[reprint]{revtex4-1}
\usepackage{graphicx}% Include figure files
\usepackage{dcolumn}% Align table columns on decimal point
\usepackage{subcaption}
\usepackage{bm}% bold math
\usepackage[usenames,dvipsnames,svgnames,table]{xcolor}
\usepackage{natbib}
\def\deg2{\hbox{$.\!\!^\circ$}}

\def\pasp{PASP}
\def\apj{ApJ}
\def\apjl{ApJL}
\def\apjs{ApJS}
\def\aj{AJ}

\def\aap{Astron. \& Astrophys.}

\def\apj{ApJ}
\def\apjl{ApJL}
\def\nat{Nature}

\def\mnras{MNRAS}
\def\aj{AJ}
\def\apjs{ApJS}
\def\aa{A\&A}

\def\pasa{PASA}
\def\aaps{A\&AS}% http://ukads.nottingham.ac.uk/abs/1981A\%26AS...45..367K
\def\aap{A\&A} %http://ukads.nottingham.ac.uk/abs/1995A\%26A...298..661B
\definecolor{mygreen}{RGB}{0, 200, 0}
\definecolor{mygrey}{gray}{0.6}

\begin{document}

%\preprint{APS/123-QED}

%\title{The hundred flavors of photometric redshifts}% Force line breaks with \\
\title{The hundred flavours of photometric redshifts}
%\title{The past, present and future of Photometric Redshifts}% Force line breaks with \\
%\thanks{}

\author{Mara Salvato}
 \altaffiliation[]{MPE}%Lines break automatically or can be forced with \\
  \email{mara@mpe.mpg.de}
 \author{Olivier Ilbert}
 \altaffiliation[]{LAM}%Lines break automatically or can be forced with \\
 \email{olivier.ilbert@lam.fr}
 \author{Ben Hoyle}
 \altaffiliation[ ]{USM,MPE}%Lines break automatically or can be forced with \\
 \email{benhoyle1212@gmail.com}


%\affiliation{
%\affiliation{
%\affiliation{%


\date{\today}% It is always \today, today,

\begin{abstract}
{\bf For more that seventy years, the measurements of fluxes of galaxies at different wavelengths and derived colours have been used to estimate their corresponding cosmological distances. 
From the fields of galaxy and AGN evolution to precision cosmology, the number of scientific projects relying on such distance measurements, called photometric redshifts, have exploded. The benefits of photometric redshifts is that all sources detected in photometric images can have some distance estimates relatively cheaply. The major drawback is that these cheap estimates have a low precision when compared with the resource-expensive spectroscopy. The methodology to estimate redshifts has been through several major revolutions throughout the last decades, triggered by increasingly more stringent requirements on the photometric redshift accuracy. Here, we review the various techniques to obtain photometric redshifts, from template-fitting to machine learning and hybrid systems. We also describe the state-of-the-art results on current extra-galactic samples and explain how survey strategy choices impact redshift accuracy. We close the review with a description of the photometric redshifts efforts planned for upcoming wide field surveys, which will collect data on billions of galaxies, aiming to solve the most exciting cosmological and astrophysical questions of today.}
\end{abstract}

%\keywords{Suggested keywords}%Use show keys class option if keyword
\maketitle

%\tableofcontents

\section{\label{sec:Intro} Introduction}
The distance of an extra-galactic source needs to be calculated before any more meaningful physical quantities can be inferred. The primary observable to measure a distance of an object is its electromagnetic spectral energy distribution (hereafter SED), which is composed of a continuum and emission/absorption lines.
The expansion of the Universe stretches the SED toward longer wavelengths by a factor $1+z$ with $z$ being the redshift. The observational difficulty in distance estimation consists of identifying a pair of characteristic features in the SED and measure the amount they have been stretched. Then, the  measured redshift is related to a proper distance, assuming a cosmological model. Emission and absorption lines are  sharp features which can be easily identified in the SED. In addition, two well known features shape the SED continuum (see Fig.~\ref{fig:photoz_principle}): 1) the {\it Balmer break} below 4000\AA{}, which is explained by the absorption of photons more energetic than the Balmer limit at 3646\AA{} and the combination of numerous absorption lines by ionised metals in stellar atmospheres; 2) the {\it Lyman break} below 1216\AA{}, explained by the absorption of light below the Lyman limit at 912\AA{} and the absorption by the intergalactic medium along the line of sight.

When SEDs of sufficient wavelength resolution are available (i.e. spectra), the emission/absorption lines can be identified and the redshift precision can be measured to better than $10^{-3}$ for a resolving power $R=\lambda/\Delta\lambda>200$ \citep[e.g.,][]{Le-Fevre:2005aa}. Despite the efficiency of the current generation of multi-object spectrographs (MOS; e.g., VLT/VIMOS, KECK/DEIMOS, SUBARU/FMOS) we can only obtain meaningful spectra for a few percent of the sources detected in deep imaging surveys even on 8-m class telescopes  \citep[e.g.,][]{Le-Fevre:2005aa}. A minimum of two well identified spectral features are required to obtain a robust redshift measurement. Given the MOS limited spectral coverage and the limited signal-to-noise in spectra for faint objects, the success rate of measuring spectroscopic redshifts (hereafter spec-z) 
can be lower than 50-70\% in deep spectroscopic surveys \citep[e.g.,][]{Newman:2015aa}.

Alternatively, by measuring the flux of an object in broader filters, we can obtain a sparse sampling of the SED sufficient to constrain the continuum shape, define the extra-galactic nature of the sources and estimate the redshift based on broad features like the Lyman and Balmer breaks, or strong emission and absorption lines. Such a low resolution distance estimate is called a "photometric  redshift" (hereafter photo-z). This principle was first applied by \cite{Baum:1957aa} to measure the photo-z of elliptical galaxies at $z \sim 0.4$. A modern version of the photo-z method has been published by \cite{Puschell:1982aa} who was the first to apply a template-fitting procedure to radio galaxies. 

The main advantage of using photo-z is that we can derive a distance measurement for all sources identified in an imaging survey. The price paid is the lower redshift precision which is typically a factor 10-100 times worse than that obtained with a low resolution spectrograph \citep[e.g.,][]{Ilbert:2009aa}. Any photo-z-based study needs an accurate assessment of the photo-z performance, which depends on the image properties (e.g. depth, wavelength coverage) and whether the sources are galaxies, active galaxy nuclei (AGN) or stars. \textit{Assessing photo-z performance requires deep and representative spectroscopic sample, which highlights the complementarity of photometric and spectroscopic redshift surveys}.

The photo-z method has been adopted as a common tool to estimate galaxy distances, and a simple ADS/NASA search shows that the fraction of refereed publications including the term "photometric redshift" has increased by a factor ten in the last two decades. This can be partly explained by the increasing number of multi-wavelength surveys. In addition, the confidence on the method has grown thanks to reassuring good comparison with spec-z's for an always increasing number of galaxies as faint as the limit of the imaging surveys. Finally, several mature and well-tested photo-z codes are both publicly available and well documented, making the technique readily accessible.

The quality of the photo-z defines the range of their possible scientific applications. Photo-z are used to study the formation and evolution of galaxies, allowing a statistical analysis of larger samples than those allowed by spec-z, also in those regions of the colour-magnitude space difficult to populate with spectroscopy\footnotetext{For example the range 1.4$<$z$<$2 is becoming accessible to spectroscopy only now that near-infrared MOS (e.g. KECK/MOSFIRE) are available}. Some of the recent applications heavily relying on accurate photo-z are: the cosmic time evolution of galaxy properties \citep[e.g.,][]{Fontana:2000aa} or the search of primordial galaxies \citep[e.g.,][]{Dunlop:2012aa}; the study of the relation between galaxy properties and their dark matter halos \citep[e.g.,][]{Coupon:2015aa} or cluster identifications \citep[e.g.,][]{Finoguenov:2007aa}. Galaxy environment or galaxy pairs evolution can also be investigated with photo-z \citep[e.g.,][]{Lopez-Sanjuan:2012aa, Man:2016aa, Etherington:2017aa}, although with a limited  precision \citep[e.g.,][]{Etherington:2015aa, Malavasi:2016aa}. Similarly, photo-z are used for studying the evolution of galaxies hosting an Active Galactic Nucleus (AGN) \citep[e.g.,][]{Miyaji:2015aa} or more peculiar objects \citep[e.g., Blazars:][]{Padovani:2012aa}.

Photo-z are also becoming a major tool to study properties of dark energy. The weak lensing tomography approach \citep[][]{Hu:1999aa} has become one of the main cosmological probes in current and future cosmological experiments such as the Dark Energy Survey (DES), High Suprime Camera survey (HSC), Euclid, and the Large synoptic survey telescope (LSST) (see \ref{sec:future}). For these applications, the required photo-z precision is not as stringent as for studying galaxy evolution, but several other challenges need to be faced. Among those are the amount of information to be processed; the need for homogeneous photo-z performances over thousands of deg$^2$; and the need to characterise the redshift distribution of a population with extreme precision.

In this review we provide an overview on the ground principles of the photo-z technique, with details on the most common methods and with an eye on current and future surveys and scientific application challenging the current state-of-art.

\section{\label{sec:method} Methods for estimating photo-z}
\begin{figure*}
\centering
\includegraphics[width=14.5cm,clip=true]{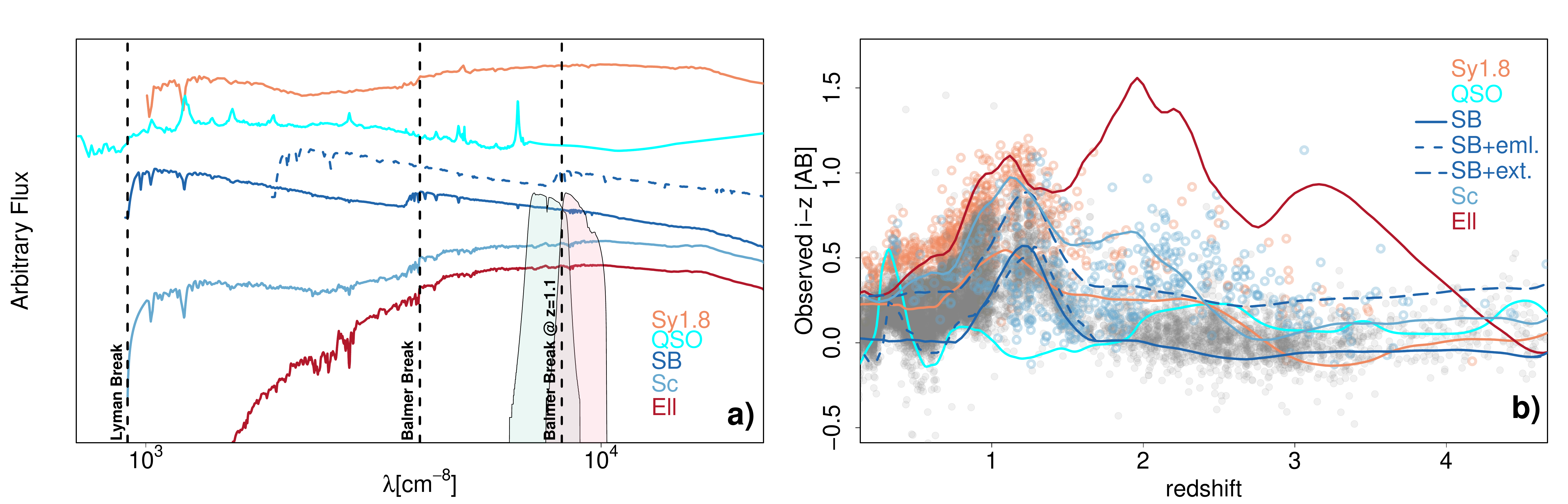}
\caption{{\bf The colour-redshift mapping principle. a):} Examples of SEDs for various type of galaxies (elliptical:Ell; Starburst:SB; spiral with small bulge: Sc) and AGN (luminous quasar: QSO; low luminosity, obscured AGN: Sy1.8 (Seyfert 1.8)) 
The Lyman and Balmer breaks, among the key features in determining the redshifts, at rest-frame  are indicated by vertical dashed lines. One template and the Balmer break are also plotted at redshift 1.1.
For clarity, the transmission curves of $i$ and $z$ filters, covering the wavelength rage between 700 and 1100 nm are also indicated.
{\bf b):} ($i-z$) color as a function of redshift. Galaxies with reliable spectroscopic redshifts are represented with grey dots while AGN obscured/un-obscured by dust are represented with red/blue small dots. The solid lines represent the expected redshift evolution of the ($i-z$) color for the templates presented in the left panel, without any extinction. The star-burst galaxy (in blue) is also shown i)considering extinction (long-dashed line) and ii)considering the contribution from emission lines (short-dashed line). 
\label{fig:photoz_principle}}
\end{figure*}

We use Fig.\ref{fig:photoz_principle} to introduce how the photo-z problem is tractable. In panel a), we show key features such as the Lyman or Balmer breaks in galaxy SEDs can be diverse.
The photo-z technique relies on the capacity to isolate the wavelength position of these red-shifted features. The breaks correspond to a rapid increase of the flux continuum from the blue to the red part of the SED. The photo-z technique aims to detect gradients between observed fluxes in adjacent filters, which would reveal the presence of a break. Panel b) of  Fig.\ref{fig:photoz_principle} illustrates this key principle. The $i-z$ color varies with redshift and reaches a maximum at $z=1.1$ when the Balmer break falls between the $i$ and $z$ bands. \textit{Thus, the first rule when designing a photometric survey is  that the filter set should be chosen to encompass key features at the redshift range of interest.}

However, the same $i-z$ color could correspond to multiple redshifts, implying degeneracies in the redshift solution. Such degeneracies can be broken by combining several colours. \textit{A second important rule is that the multi-wavelength coverage should be as broad as possible to limit the risk of photo-z degeneracies} \citep[e.g.][]{Benitez:2009aa}. 

\begin{figure*}
\centering
\includegraphics[width=14cm, angle=0]{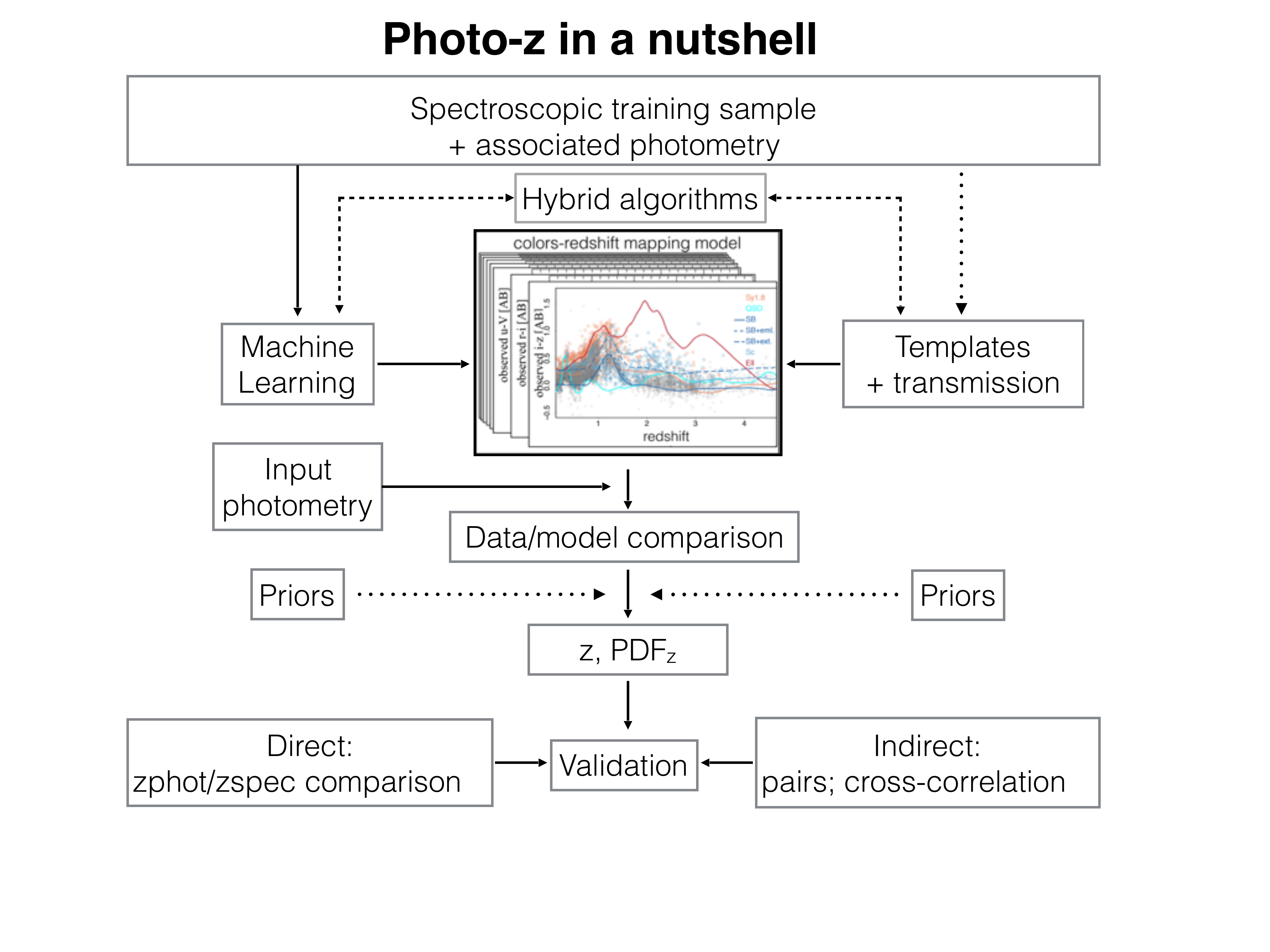}
\caption{{\bf Photometric redshift technique in a nutshell.} Flowchart representing the various steps involved in computing photometric redshift.The core of ML, template-fitting and hybrid methods is the multi-dimensional colour-redshift relation. The spectroscopic sample, essential for ML, is optional (although desirable for some tuning) for template-fitting techniques.  With the accumulated knowledge over the past decade we are able to define reliable priors that are improving dramatically our results. After the computation of photo-z and PDF, a huge effort goes into the validation of the results. \label{fig:method}}
\end{figure*}

The flow-chart of  Fig.~\ref{fig:method} shows the fundamental steps and ingredients used in all photo-z techniques. At the core there is always a model of the mapping between various colours (or fluxes) and redshift. The redshift solution and its associated Probability Distribution Function (PDF) are obtained by comparing this mapping and the observed fluxes of the studied source. In the case of template-fitting methods, the redshift-colours mapping is based on physical knowledge built by scientists over time. With Machine Learning methods, the mapping is obtained every time anew, using a representative `training sample' of galaxies with both photometry and known redshifts. For both methods the results can be improved by using additional priors (see further in this section for more details).

While the basic principle is simple, the mapping and its application to the data  can be established in many different ways. In this section, we highlight the most commonly used methods.

%----------------------------
%----------------------------
%----------------------------
%----------------------------

\subsection{Physically motivated methods}

In this branch of photo-z techniques, the mapping between flux and redshift is predicted for extragalactic sources taking into account  physical processes regulating the observed light emission.

Thus, the most basic ingredient is the definition of a set of SED templates, which could be either obtained from theory or observations. Theoretical templates are generated using stellar population synthesis models \citep[e.g.,][]{Fioc:1996aa, Bruzual:2003aa, Maraston:2005aa, Conroy:2012aa} which progressed significantly in the last decades. Such theoretical templates rely on numerous assumptions, for instance on the galaxy star formation histories. Empirical templates are extracted from observed spectra \citep[e.g.,][]{Coleman:1980aa, Kinney:1996aa}, extended over the entire wavelength range with models \citep[e.g.,][]{Noll:2004aa, Polletta:2007aa}. Not only the type of templates determines the quality of photo-z, but also their optimal coverage of the colour-redshift space \citep[e.g.,][]{Chevallard:2017aa}.

Nebular emission lines emitted by HII regions should also be considered in the templates \citep[e.g.][]{Ilbert:2006aa, Schaerer:2012aa, Pacifici:2012aa} .
\cite{Ilbert:2009aa} have clearly shown how considering the contribution of the nebular emission improve on the accuracy of photo-z by a factor of 2.5.  The importance of this component is  easily understood looking at the panel b) of  Fig.~\ref{fig:photoz_principle}. There, the SED of a starburst galaxy is plotted with and without a contribution from emission lines, with the former better matching the colour-redshift map of spectroscopically confirmed galaxies. 

To obtain accurate photo-z at $z>1$ it is particularly important to model the effect of dust attenuation, because optical data will sample the rest-frame UV part of the SED which is most affected. The dust grains present in the interstellar medium absorb and scatter the stellar light, and reddens the SED continuum. Depending on the amount of dust and its grain properties, the impact can be severe (see panel b of Fig.~\ref{fig:photoz_principle}, comparing long-dashed and continuous lines).  When computing photo-z with template-fitting codes the dust attenuation is typically modelled as a free parameter, using one of several dust attenuation laws \citep[e.g.,][]{Calzetti:2000aa, Prevot:1984aa}. 

Then, the templates are corrected for the absorption of the galaxy light that get absorbed when crossing the neutral gas present in the inter-galactic medium \citep[e.g.][]{Madau:1995aa, Draine:2011aa}.

The light is also attenuated by the dust from our Milky Way along the optical path between the  source and the observer. Normally, catalogues are corrected for these effects using the dust extinction maps of \cite{Schlegel:1998aa}, but a better treatment is proposed in \cite{Galametz:2017aa} which accounts for variations in the intrinsic SED of galaxies.

Finally, the redshifted templates should be integrated through the filter transmission curves to produce the modelled fluxes. All other factors that could modify the light distribution should be considered, such as the impact of Earth atmosphere, the optics of the telescope, the shape and efficiency of the filter curves, the Charged Coupled Device (CCD) transmission. These various components are usually integrated into one single transmission curve per filter and stored within the code libraries. 

As a final note, while this section discussed the modelling of extra-galactic sources, template-fitting techniques also need to consider stellar templates, because we can not rule out the Galactic nature of the studied sources. Typically, the fit to the stellar templates is performed independently and the extragalactic/Galactic nature of the source is decided a posteriori.

\subsection{\label{MachineLearning} Data driven methods}

Rather than relying upon any physical models one may estimate photo-z using data driven approaches. The most common method is Machine Learning (hereafter ML). 
ML algorithms can be  categorised into unsupervised and supervised learning, the difference being that unsupervised learning requires only photometry, while supervised methods require both photometry and reliable spec-z for a data sample during the training of the algorithm. In the photo-z framework, these training samples allow the algorithm to {\it learn} the mapping between colour and redshift (see e.g, Fig.\ref{fig:photoz_principle}).  

In general all supervised ML methods perform function approximation, and for photo-z this entails finding a function which maps between the multi-dimensional photometry space (e.g., fluxes or colours) and the redshift values of the training sample. Indeed the function is assumed to be ``well behaved'' between the multi-dimensional locations of the training data, and thus supervised ML methods can be viewed as performing high dimensional interpolation.  Once this function has been learned, the photometry of a source with an unknown redshift is localised in the multi-dimensional photometry space and paired with a corresponding redshift value of probability distribution.
Due to the interpolative nature of supervised ML approaches, one must ensure that the space of photometric properties of the sample for which predictions will be made, is well sampled by the training examples \citep[see, e.g.,][]{Hildebrandt:2010aa, Cavuoti:2012aa, Beck:2017aa}. Failing this, the algorithms may perform extrapolation and lose accuracy. Furthermore, the complexity of the (unknown) manifold to be approximated using ML provides an estimate of the number density of training data in each region of color space that are required. 

Two popular supervised  ML systems are often used for photo-z computation, Random Forests \citep[][]{Breiman:2001aa} and Neural Networks \citep[][]{Rosenblatt:2005aa} which we briefly describe below. 

Random forests \citep[e.g.,][]{Carliles:2010aa, Carrasco-kind:2013aa} are aptly named because they are constructed from collections of `decision trees' and each tree is grown using a random subset of the training data. Decision trees are a set of logical if-statements that group the properties of the training sample into cells, which are defined to minimise the spec-z dispersion of the data in each cell.
Successively, each cell is assigned the average value of the spec-z. Once all the trees have been grown, predictions are made on the redshift of the sources in a test sample  by passing their photometric properties through each tree and averaging over all of the obtained redshifts. The number of trees or the maximum number of cells are some of the variables of the algorithm.

Neural Networks are a class of popular ML algorithms that were inspired by how neurons in the animal brain process information and learn.
The NN algorithms \citep[e.g.,][]{Collister:2004aa, Vanzella:2004aa, Bonnett:2015aa,Cavuoti:2012aa,Brescia:2015aa} make increasingly complex non-linear matrix transformations of the input properties. The structure consists of different layers, and each layer can be viewed as a new transformation of those which come before it. The output of the NN is defined by the user, and represents the target value of interest, such as the redshift value. The NN algorithm tunes the matrix transformations using the training data, such that the output value minimises the residuals between true redshift and predicted redshift.

Finally, Deep Machine Learning (DML), the current state of the art in computer science,  has been also applied to problem of photo-z estimation in \citep[e.g.][]{Hoyle:2016aa, Disanto:2018aa}. DML is based on normal neural networks, but with many thousands of neurons in each hidden layer. In particular DML may also work directly from  galaxy images. On many square degrees surveys this means avoiding the alternative of computing photo-z gathering photometry from catalogs of different surveys, often using different measures.

Given that supervised ML methods require training data with both photometry and spectroscopic redshifts to approximate the mapping between colours and redshift, they are of course limited by the nature of the training data, often limited to low redshift and bright luminosity (see Fig.~\ref{fig:accuracy}). Given the dearth of both high redshift (e.g., $z>1$) and very faint galaxies, machine learning methods should be used with caution in these regimes.

Unlike supervised ML, unsupervised ML algorithms does not use spec-z in the training phase, but rather performs clustering in the input data space to identify groups of similar objects. The most popular example in the photo-z literature is the Self Organising Map (SOM). A nice introduction to SOMs can be found in \citep[e.g.,][]{Carrasco-Kind:2014aa, Masters:2015aa}. 

All ML algorithms used for photo-z are able to generate PDFs of varying sophistication. The simplest method is to randomly sampling from values and errors of each input parameter many times, and associate the normalised distribution of predictions as a PDF \citep[e.g.,][]{Carrasco-kind:2013aa, Cavuoti:2017aa}.

Other exploratory techniques from ML have also been ported to photo-z estimation, such as anomaly detection, data augmentation, and feature importance \citep{Hoyle:2015aa,Hoyle:2015ab,Hoyle:2015ac}. In turn, these techniques allow ML methods to identify problematic training data, provide some extrapolation ability outside of the original training sample, and motivate the choice of input parameters that provide the most photo-z predictive power. As in the case of template-fitting methods, ML can also improve on the results by considering additional information as a prior, often in the form of re-weighting the training data \citep[][]{Lima:2008aa}.

\subsection{\label{subsec:priors} Using additional data as prior information.} 

Many photo-z codes, often produce a redshift $z$ which is the maximum of the likelihood $\mathcal{L}(z | T, D)$ given the data $D$ and galaxy type $T$ or choice of SED. This redshift inference can be improved by adding other prior information based on the empirical knowledge that we have acquired over time of the redshift evolution of various galaxy properties.

When the prior is considered in a Bayesian context, the posterior redshift prediction is the product of the multi-dimensional likelihood and the prior which is then marginalised over, to produce a one dimensional posterior in redshift. 
Such a Bayesian approach has been established in BPZ \citep[][]{Benitez:2000aa} by introducing a prior on the redshift distribution  per galaxy spectral type and magnitude bin. Other choices of prior have also been recently made, e.g. as luminosity function for the GOODZ code \citep{Dahlen:2010aa} and  mass-SFR relation in \cite{Tanaka:2015aa}. One of the most recent public code is  BEAGLE \citep{Chevallard:2017aa}, which is among the few that adopt a fully Bayesian approach and thus a prior can be assigned to each parameter.
Prior should be used with care given the risk of introducing an inaccurate information (e.g. the redshift distribution  beyond the spectroscopic limits is hardly known). Still, in the case of sparse wavelength coverage which produce several redshift peaks in the PDFs, \cite[][]{Benitez:2000aa} showed that using the most probable solution based on prior knowledge helps in reducing the fraction of catastrophic failures for the average population, at a price of associating the wrong photo-z to specific populations (as high-z sources, or AGN). Independently on the type of prior, we believe it is a responsibility of the authors to describe in detail the priors that are adopted and to discuss the implication that the prior may have on the photo-z results.

\subsection{\label{subsec:spatial} Introducing spatial information}

The correlation of source positions can also be used as a redshift estimation technique. This method uses the fact that galaxies are not distributed randomly in the Universe but reside in large scale structures. By using the spec-z of a reference sample, the redshift distribution of an unknown sample can be found by maximising the spatial cross-correlation signal between the unknown and the reference samples. As early as \cite{Seldner:1979aa}, galaxy clustering was used to extract information on QSO distances. This method has been mainly revived in the last decade to estimate the mean redshift of photo-z selected samples for weak lensing applications \citep[e.g.][]{Newman:2008aa}. While considering usually only the large-scale clustering, \cite{Menard:2013aa} generalise the method using also clustering on smaller scales. However, such approach requires the unknown sample to be preselected in a narrow redshift range, in photo-z or in colours \citep[e.g.][]{Scottez:2016aa}. By considering extremely narrow sample of preselected galaxies, it becomes even possible to measure individual redshifts \citep[][]{Rahman:2015aa}.

\cite{Aragon-Calvo:2015aa} propose a new estimate of the photo-z using the cosmic web. Such method requires a dense spectroscopic coverage to characterise accurately void regions, filaments, walls and clusters. The redshift is obtained from the product between the photo-z derived from the colours, the cosmic web and the density field. One of the limitation of the method is the need of using accurate photo-z (better than 1\%) from colours in order to avoid a catastrophic association between the photo-z and the wrong cosmic web structure.  For 3\% of the SDSS multi-color catalogue, \cite{Aragon-Calvo:2015aa} were able to apply this method, and show that an accuracy as good as $10^{-5}-10^{-4}$ was possible.

\section{\label{sec:comparison} Which method for which survey}
But how to decide between different methods?
 Is template-fitting better than ML? Or vice versa?
 Or, is there among the SED-fitting and ML codes a favoured one?
In general, a direct and fair comparison between performances of different template-fitting and ML codes is possible only within tests specifically designed for it \citep[e.g.,][]{Sanchez:2014aa, Hildebrandt:2010aa, Dahlen:2013aa, Beck:2017aa, Duncan:2017aa}. Here we list the most used and/or public codes, together with their key features, pointing to the papers that attempted a more quantitative comparison of the various algorithms. More on this is also discussed in Sec.~\ref{sec:future}.
 
\subsection{\label{SFcomparison} Template-fitting codes}
Numerous template-fitting codes are publicly available, with Hyperz \citep[][]{Bolzonella:2000aa} being a precursor. While the ultimate common goal of all the codes is to compute photo-z, each code developed their own specificity. 
For example, template libraries change from code to code. EAZY \citep[][]{Brammer:2008aa} combines basic templates and creates new ones on-the-fly, while ZEBRA \citep[][]{Feldmann:2006aa} trains the templates using a spectroscopic sample (a risk, when the spectroscopic sample is not representative of the entire population \citep[see Fig.11 and related text in ][]{Luo:2010aa}). Some codes include dust attenuation in their templates \citep[e.g. BPZ, ][]{Benitez:2000aa} while others include dust as a free parameter (e.g. Hyperz). No papers have yet to clearly highlight that one template set is superior to another. Still, EAZY include an error function associated to the templates which improve the quality of the PDF.

Depending on the dataset and the scientific objective, some features are crucial to improve the photo-z accuracy. It is now widely accepted that galaxy photo-z performances improves, especially at high redshift, when including emission lines \citep[][]{Schaerer:2012aa} in the templates, like in Le Phare \citep[][]{Arnouts:1999aa,Ilbert:2006aa} or EAZY, especially when the photometry from narrow and intermediate  bands are used like in COSMOS \citep[][]{Ilbert:2009aa,Hsu:2014aa}, MUSYC \citep{Cardamone:2010aa}, SHARDS \citep[][]{Perez-Gonzalez:2013aa} and ALHAMBRA \citep{Molino:2014aa}.

Another example of adapting the code configuration to the photometric data available is the use of prior. \cite{Tanaka:2015aa} showed that for a faint survey at $i<25$ and with only \textit{griz} bands, he was able to decrease the fraction of outliers from 45\% to 25\%. But if the survey provides an extensive  multi-wavelength coverage including NIR, adding a prior is usually not considered, meaning that an equal probability is associated to each value of the parameter. Another important aspect is the sensitivity of the photo-z bias to the absolute flux calibration of a survey: \cite[][]{Ilbert:2006aa} showed that only a few percent uncertainty in the absolute calibration of the photometry could create a bias at the same order as the photo-z precision for even bright sources. This calibration issue can be corrected using a spec-z training sample, with a overall improvement in accuracy by a factor of 2. The zero-point calibration was also pointed out in \cite[][]{Dahlen:2013aa}, as a way to reduce the fraction of outliers and to increase the accuracy.

\subsection{\label{MLcomparison} ML codes}

The choice about which ML algorithm will provide the best performance for a particular dataset or problem is nontrivial. Results from computer science show that if the training sample is large enough, and the training time of each algorithm is long enough, then the performance from different algorithms converges \citep[see e.g.,][for a photo-z setting]{Hoyle:2015ab,Beck:2017aa}. If the training samples are small, or the training times are limited, then the performance of different codes depends on the complexity of the underlying surface to be learned. Given the plethora of ML algorithms on the market, each group typically considers only a few algorithms for any given problem, often chosen from those of which they have obtained good results in the past. Furthermore those algorithms which are finally implemented in large scale astronomical surveys, such as the TPZ \citep[][]{Carrasco-kind:2013aa}, Skynet \citep[][]{Bonnett:2015aa} and aNNz2 \citep[][]{Sadeh:2017aa} methods for DES, are often those in which the code authors were part of the collaboration, and had a vested interested in seeing their codes advanced. A good starting point for deciding which ML algorithms should be applied to new datasets, is to use those which are very fast to implement so that any subsequent performances can be easily bench-marked. For this purpose the authors would suggest starting with decision tree based algorithms, such as Random Forests.

\subsection{SED-fitting or ML?}

Generally, the choice between template-fitting and ML methods strongly depends on the scientific application and on the spec-z sample available for the training. For instance, ML applied to SDSS data outperform template-fitting for low-redshift sources. This is because the relatively few optical bands are still sufficient to accurately correlate with redshift, and the spectroscopic sample used for the training is rich and complete. But in deep pencil-beam surveys covering a large redshift range, the reverse case is shown in \cite[][]{Hildebrandt:2010aa}. In other words, the use of ML is limited to those surveys with a sufficient training set, while if the scientific objective is to study galaxy population with limited spectroscopic coverage, template-fitting code should be favoured. 
 
Template-fitting codes could also model uncertainties in the absolute calibration, but ML methods are also insensitive to photometry biases depending on magnitude or colours. Therefore if the goal is to limit biases over a large field  well covered with spectroscopy, ML would normally expected to be the favoured approach. In term of speed, ML clearly outperform the template-fitting techniques \citep[][]{Vanzella:2004aa}. One of the largest benefits of ML algorithms come from their optimisation to work smoothly on massive data sets. They have been specifically developed to process massive volumes of data, and can easily accommodate current and future astronomical sized data sets \citep[see e.g.,][]{Vanzella:2004aa}.

\section{\label{sec:validation} validation and state of art}
\subsubsection{\label{subsec:perfVal} Photo-z performance and validation}

An extensive characterisation of the photo-z performance should be made for every catalog and sub-sample of interest. It is typical to compare a photo-z point prediction (such as the mean or mode of the PDF, $z_{\rm phot}$) and the spec-z ($z_{\rm spec}$) as a way to assess performance. 
Note that any spec-z which are used for ML training, or prior construction,  should generally be excluded from the validation sample.

The following three measures are often

adopted to assess photo-z performances:
\begin{itemize}
\item Bias: defined as $<z_{\rm phot}-z_{\rm spec}>$, it characterises the average separation between the predicted and the true redshift \citep[e.g.,][]{Bordoloi:2010aa}.
\item Outliers (or catastrophic failures) fraction: usually defined as the fraction of sources for which $\|z_{\rm phot}-z_{\rm spec}\>| > N \sigma$ \citep[e.g.,][]{Dahlen:2013aa}, or $|z_{\rm phot}-z_{\rm spec}|/(1+z_{\rm spec}) > 0.15$ \citep[e.g.,][]{Hildebrandt:2010aa}. It highlights the fraction of sources with unexpectedly large errors. 
\item Precision ($\sigma_{z_p}$): is often defined as the standard deviation of $(z_{\rm phot} -z_{\rm spec})/(1+z_{\rm spec})$ or as $1.48 \; \times \; {\rm median}(| z_{\rm phot} -z_{\rm spec} |/(1+z_{\rm spec}))$ \citep[][]{Ilbert:2006aa}, the latter being less sensitive to outliers. The precision describes expected scatter between predictions and truths.
\end{itemize}

These metrics are not unique \citep[e.g.. see discussion in Sec.4 in][]{Dahlen:2013aa} and they could be science-dependent. However, because they are the most used, they allow for a easy comparison between surveys.

We note that photo-z performance measured using a direct comparison between photo-z and spec-z is  less meaningful if the spectroscopic coverage is not fully representative of the entire photo-z sample, e.g., if the spec-z sample is biased toward bright sources \citep[see e.g.][]{Bonnett:2016aa, Beck:2017aa}. 

When this happens, a statistical mapping between the photometric parameter space covered by the two samples should be defined \citep[e.g.][]{Masters:2015aa}. For regions of photometric space without spectroscopic coverage, alternative methods have been developed to establish the photo-z performances. For example, the galaxy closed pairs technique \citep{Quadri:2010aa, Hsu:2014aa}, uses the fact that close pairs have a significant probability of being associated and that they therefore should have similar redshifts. The comparison of the spatial cross-correlation between two redshift bins with that expected from a model may also be used to measure redshift precision and contamination \citep[][]{Benjamin:2010aa}.

Characterising the performance of the full photo-z PDF is slowly becoming popular, as it can highlight the lack of templates, training sample, or sub-optimal redshift prediction routines. One usual problem affecting PDFs is caused by under-estimated photometric uncertainties, creating a PDF peak which is too narrow \citep[e.g.][]{Dahlen:2013aa}. In numerous works authors validate the PDFs by insuring that $68\%$ of the spec-z falls within the 1$\sigma$ uncertainties derived from the PDF. In \cite{Bordoloi:2010aa} the authors introduce the statistical method of the Continuous Rank Probability Score (CRPS) to the photo-z community as a means of testing PDF performance. The CRPS asserts that the value of the cumulative PDF at $z_{\rm spec-z}$ must be uniform for an ensemble of galaxies.

\subsubsection{\label{subsec:state} State of the art}
\begin{figure*}

\centering
\includegraphics[width=15.cm,clip=true]{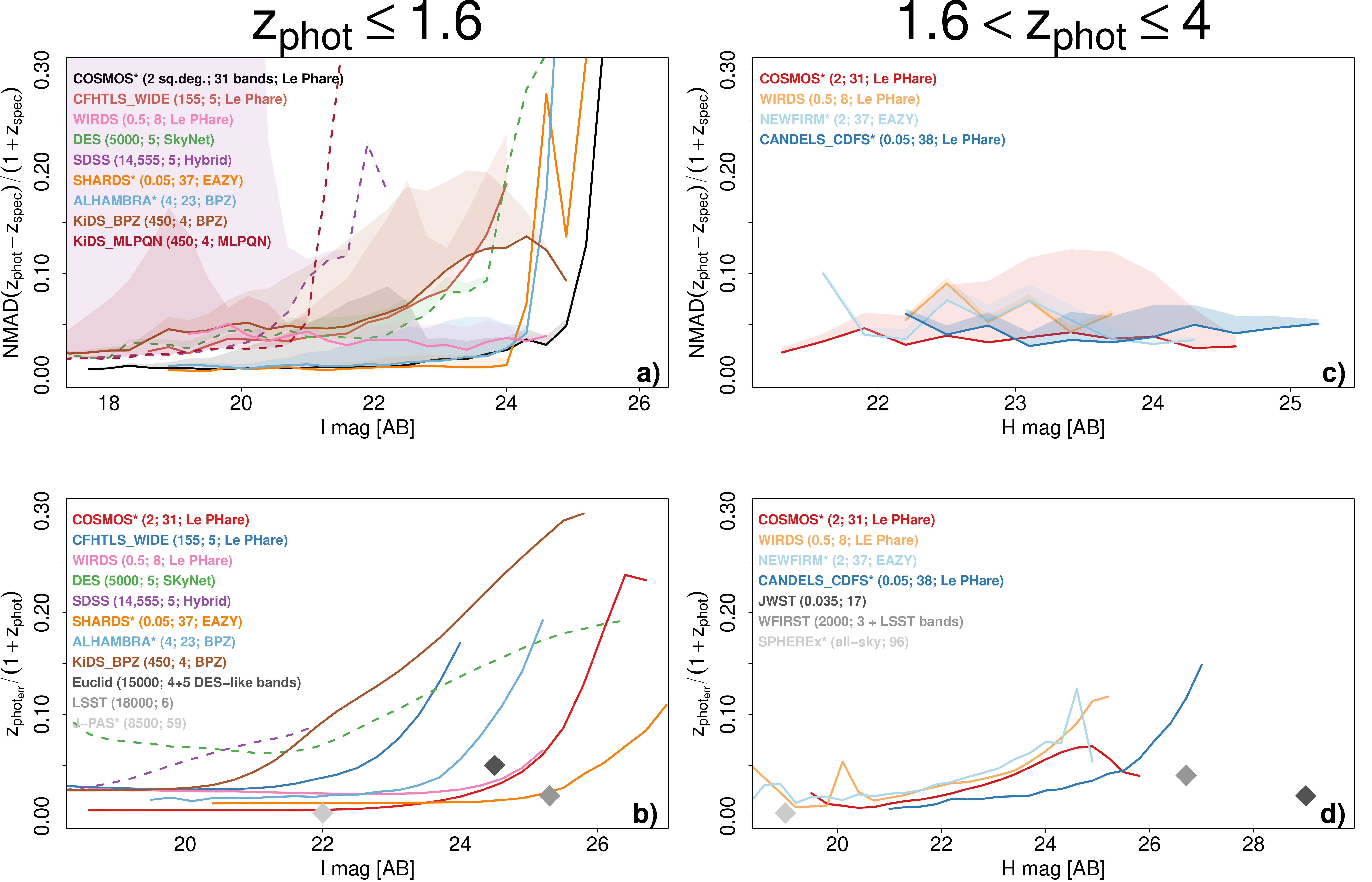}
\caption {\small {\bf Accuracy of various photo-z surveys as a function of their depth and number of bands}. The low redshift range is shown in the panels a) and b)  where the accuracy is measured as function of the $i$ band. For the high redshift range (panels c) and d)), the accuracy is measured as a function of the near-infrared depth in the $H$ band. In all panels only sources with a photometric error smaller than 0.3 are considered. The accuracy is measured with NMAD, with the solid lines indicating the surveys where photo-z were computed via template-fitting, while the dashed lines indicate the surveys with photo-z computed via ML or Hybrid methods. Surveys with intermediate or narrow band photometry (COSMOS: \citep{Laigle:2016aa}; SHARDS: \citep{Perez-Gonzalez:2013aa}; ALHAMBRA: \citep{Molino:2014aa}; CANDELS/CDFS: \citep{Hsu:2014aa}; NEWFIRM: \citep{Whitaker:2011aa}) are highlighted with an asterisk. The coloured areas are proportional to the number of reliable spectroscopic redshifts used for the training, at that magnitude, while the numbers in brackets beside each survey indicate their size in square degrees, the number of bands and the code used}. In general, the less accurate photo-z  are at the faint end of each survey.
A measure of the accuracy beyond the availability of spectroscopic redshift is provided in panels b) and d), where the solid lines show the median photo-z error divided by (1+z$_{phot}$), as a function of the depth of the survey.
It is clear how the accuracy in all the panels is better in the region well populated by the spectroscopy, where the sources are brighter.
Surveys with intermediate and narrow band photometry perform better than those having only broad band photometry (WIRDS: \cite{Bielby:2012aa}; CFHTLS\_WIDE: \cite{Coupon:2009aa}; KiDS: \cite{DeJong:2017aa}; SDSS:\cite{Beck:2016aa}; DES: \cite{Bonnett:2016aa}). Finally, with grey symbols we add the expected accuracy for the future cosmological surveys.
\label{fig:accuracy}
\end{figure*}

The wavelength coverage (see panel a) of Fig.~\ref{fig:degeneracy}) and the quality of the input data determines the accuracy of the photo-z, independently of the merits of each redshift code. One of the main advantages of photo-z is the ability to estimate distance information for faint sources, granted that the photometric errors associated to the measures are sufficiently small to constrain the redshift by limiting the degeneracy in the solutions. Hence, because the photometric errors are larger when the sources in an image are faint, the only way to keep small the photometric errors (and thus keep high the accuracy of photo-z) is to obtain images  as deep as possible (see Fig.\ref{fig:accuracy}). The photo-z estimates in CANDELS represent one of the deepest photo-z samples available today with $\sigma_{z_p}$ increasing from 0.040 (8\%) to 0.055 (28\%) between $H<24$ and $26<H<28$, respectively. Over large fields of several deg$^2$, $\sigma_{z_p}\sim 0.05$ is routinely reached at $i<26$ \citep[e.g.][]{Laigle:2016aa}. 

With broad band photometric data alone, the best $\sigma_{z_p}$ is limited to approximately 0.025 (see Fig.\ref{fig:accuracy} for results on pencil-beam and wide area surveys) and this does not seem to improve even above large signal-to-noise values \citep[e.g., SN$>$40;][]{Coupon:2009aa}. In the last decade, medium band data with filter widths around 400\AA{} have enabled a breakthrough in precision by improving the SED resolution. COMBO-17 \citep{Wolf:2004aa} was the first survey to use medium bands imaging to produce a photo-z catalogue. There, $\sigma_{z_p}$ of bright sources ($i<22$) reached $\sim 0.02$, thanks also to the ability to precisely locate the Balmer break. The COSMOS, SHARDS and ALHAMBRA surveys improved upon this precision to $\sim 0.01$ out to a redshift $\sim$1.5 with deeper medium-band photometry and by introducing emission lines into the templates.

The increase in sensitivity of NIR detectors has allowed astronomers to extend their study beyond the \textit{redshift desert} by following both the Balmer break between $1.6<z<4$ \citep[e.g.][]{Cimatti:2002aa}, and the Lyman break above $z>8$ \citep{Bouwens:2015aa}. In this way high precision photo-z ($\sigma_{z_p}\sim 0.02$) has been reached  between $1.6<z<4$ for massive galaxies in the NEWFIRM survey \citep{Whitaker:2011aa}. 

Photo-z enable the identification of galaxy samples across thousands of deg$^2$. For example the DES survey is reaching depths of $i\sim 24$ over 5000 deg$^2$, with a photo-z accuracy expected to be below $\sim 0.08$. The major challenge for wide-field surveys is the calibration of photometric noise precision over large areas while being observed over time scales of many years with varying sky conditions, and a potential degradation in quality of the instruments. 
Also, the spectroscopic training sample should not be limited to a small area of the sky, but rather homogeneously distributed.

Finally, beside these other difficulties, the procedure to extract the photometry from images is also crucial. Source extraction is commonly performed with SExtractor \citep{Bertin:1996aa} but the region within which the galaxy flux is measured is important because too large an area would compromise the signal-to-noise of faint sources \citep{Hildebrandt:2012aa, Moutard:2016aa}. Moreover the light should be produced by the same regions of the studied sources despite the Point Spread Function (PFS) variation from one band to another. In cases of PFS variation between the various bands, PFS homogenisation is necessary \citep[e.g.][]{Grazian:2007aa, Ilbert:2009aa, Hildebrandt:2012aa, Laigle:2016aa}, and efficient tools are being developed to produce fluxes measured in a consistent way using the high-resolution images as reference for example see PYGFIT : \citep{Mancone:2013aa}; Synmag: \citep{Bundy:2012aa}; TFIT: \citep{Laidler:2007aa}; T-PHOT: \citep{Merlin:2015aa}. 

%-------------------------------------
%-----------------------------------------

\section{\label{sec:exotics}  Photo-z of exotic sources}
\begin{figure*}
\centering
\includegraphics[width=15cm,angle=0,clip=true,trim=0 0 0 0]{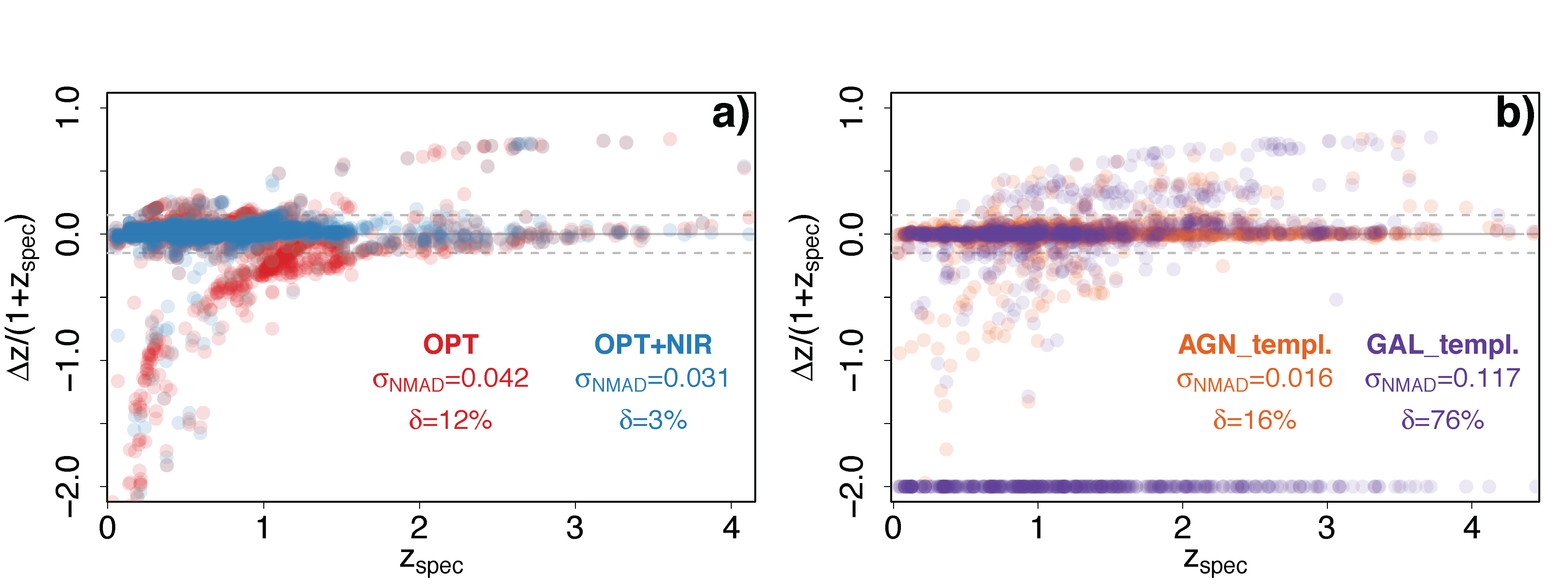}
\caption{ {\bf Impact of multi-lambda coverage and choice of templates.}\\
{\bf a):} Comparison of photo-z for galaxies computed using the optical bands (ugriz; red) or the optical and near infrared (NIR) (ugrizJHK; blue). The addition of NIR infrared photometry improves dramatically the precision at z$>$1. Sources external to the dashed lines $\pm$ 0.15*(1+z\_spec) are considered outliers.
{\bf b):} Difference between photo-z obtained using either galaxy (violet) or AGN (orange) templates for sources that are X-ray detected in {\it Chandra} Legacy COSMOS and with a reliable spec-z. The 33\% of sources for which no template of normal galaxy could compute the photo-z are artificially set to -2.
}
\label{fig:degeneracy}
\end{figure*}

The discovery that virtually every galaxy hosts a super massive black hole \citep[e.g.][]{Magorrian:1998aa}, suddenly increased the interest of the scientific community for those galaxies hosting an AGN, now seen as a key ingredient in galaxy evolution models. Therefore, computing their redshifts became crucial. However, while the  number density of AGNs is sufficiently high that spec-z follow-up via single slit spectroscopy would be too time consuming, they are so sparse that a multi-object  spectroscopic campaign would be totally inefficient.
Thus, the photo-z technique is favoured, although these sources are particularly challenging. In fact, their  SED is characterised by the sum of two unknown relative contributions, host andAGN,  and in any photometric band this depends on the type of host and the type and strength of the AGN \citep[e.g., Figure 5 in][and related text]{Bongiorno:2012aa}. When these constraints are ignored, and for example only templates of galaxies are used (i.e. assuming a dormant black hole), the photo-z obtained with SED-fitting will be characterised by a high number of degenerated photo-z solutions, or, worse, catastrophic failures. This is clearly demonstrate in the right panel of Figure~\ref{fig:degeneracy}. There, we show the difference between photo-z computed using either AGN \citep{Salvato:2011aa}  or galaxy \citep{Laigle:2016aa} templates, for a sample of 1672 X-ray selected AGN with a secure spec-z from the Legacy {\it Chandra} COSMOS field \citep[][]{Marchesi:2016aa}. In particular, the discrepancy between the photo-z solutions increases often with the strength of the AGN \citep[][]{Salvato:2011aa}.

%----------------------------
%----------------------------

The choice of templates to include in a library is dictated by the type of AGN that are treated \citep[e.g., depending on their hardness ratio][]{Zheng:2004aa}, the selection technique (e.g. excess flux, variability) used for the identification and by the selection band (e.g., X-ray, radio, optical, infrared) and its depth.
 This is clearly demonstrated in Figure 14 of \cite{Hsu:2014aa}, where the X-ray flux distribution of the AGN detected in various X-ray surveys is shown. Clearly, the population of sources that are represented in the wide but shallow XMM-COSMOS \citep[e.g.][]{Brusa:2010aa} survey, is very underrepresented in the deep but narrow {\it Chandra} observations of the Chandra Deep Field South field \citep[e.g.,][]{Luo:2010aa}. Reciprocally, the bulk of the sources in the latter field is not represented at all in the former. Similar results are shown recently in \citep{Duncan:2017aa} for Radio selected AGN. This is particular relevant considering that wide or all-sky surveys such as eROSITA \citep[][]{Merloni:2012aa} in X-ray and EMU \citep[][]{Norris:2011aa} in the Radio are expected to detect million of AGN that will be used as cosmological probes and for evolutionary studies.

The problems is alleviated in rich photometric datasets which are inclusive of both intermediate and narrow bands filters because they help to identify emission lines \citep[][]{Salvato:2009aa, Cardamone:2010aa, Hsu:2014aa}, which are the only feature visible in the power-law continuum of a QSO (bright AGN outshining their host) for which colours are independent of redshift (see for example the cyan track in panel b) of \ref{fig:photoz_principle}). 

Priors are a crucial ingredient when computing photo-z for AGN via template-fitting. 
Specifically, given the typical absolute magnitude of these sources (typically brighter than -22 in blue optical bands), a prior based on the apparent size allows to limit the range in redshift solution and thus the degeneracy \citep[e.g.,][]{Salvato:2009aa, Salvato:2011aa, Kitsionas:2005aa}.

ML techniques have also estimated photo-z for AGN but the best results are mostly reliable for QSO at high redshift, where the host contribution can be neglected. Examples of some successful applications are presented in \cite{Bovy:2012aa, Brescia:2013aa} and \cite{Budavari:2001aa}. The mentioned works are all focused on the SDSS footprints, where the plethora of spectroscopic follow-up data are well suited to ML techniques. Very recently \citep{Mountrichas:2017aa} applied ML to a sample of bright X-ray selected AGN, providing reliable results also at low-redshift. \\
An additional problem when computing photo-z for AGN is that variability across the spectrum is an intrinsic property of the sources, with different cadence and intensity. This makes it difficult to reliably estimate photo-z when multi-wavelength data are collected over time scales of years \citep{Simm:2015aa}.

Variability does not only effect QSOs, but also objects as Bl-Lac, Gamma-ray Bursts (GRBs) and Supernovae (SN).
Bl-Lac belong to the family of AGN, in particular Blazars, i.e. sources that are observed through the jet launched from the center and for this reason are characterised by a featureless power-law continuum, unusable for assessing redshift via spectroscopy. Here, simultaneous observations  in multiple bands are required in order to measure the absorption  bluer than Ly$\alpha$ by the intervening material \citep{Rau:2012aa}. Gamma-Ray-Burst (GRB) and Super Novae (SN) are sudden and quickly occurring explosive events that allow us to measure the distance of galaxies that would be otherwise too faint to be detected. For GRBs, we rely on simultaneous observations in optical+NIR and SED fitting templates \citep{Kruhler:2011aa}, while the most reliable photo-z for SN rely on the use of priors \citep[e.g.][]{Palanque-Delabrouille:2010aa}. 

\section{\label{sec:future} The Future}

\subsection{Reaching the deep Universe}

The photo-z technique gained momentum when the first images from the Hubble Deep Field became available in 1995. \cite{Lanzetta:1996aa} published a first photo-z catalogue to study 1683 galaxies out to z=6, which was an incredible leap forward in studying the high redshift Universe. Such analysis was followed by numerous attempts to make the photo-z estimation technique more robust \citep[e.g.][]{Arnouts:1999aa, Bolzonella:2000aa}, and to exploit the following generations of HST deep surveys \citep[e.g.][]{Dahlen:2013aa}. The template-fitting technique has been central since only sparsely spectroscopic coverage is possible in these deep fields. The study of primordial galaxies will trigger a new burst of activity in the next years with the launch of the James Web Space Telescope \citep[JWST]{Finkelstein:2015aa}. With the efficient near-infrared NIRCAM camera, surveys could be conducted at depths of $mag_{AB}\sim 30-31$ also by adopting a strategy similar to the CANDELS survey. \cite{Bisigello:2016aa} showed that NIRCAM is a powerful  camera to produce photo-z above $z>5$ out to $z \sim 20$. 

Based on the HST experience, template-fitting will probably remain the main photo-z technique in these extremely faint fields. However, new difficulties will arise because of the increasing contribution of the emission lines with redshift \citep{Schaerer:2010aa} and an evanescent Balmer break feature. While  emission lines create gradients in the colour-z relation that could be used to measure a redshift \citep[e.g.][]{Labbe:2013aa}, it makes also the estimate more challenging because more prone to degenerated solutions \citep{Bisigello:2016aa}.  The knowledge of the galaxies and space at redshift beyond $z>10$ will be soon revealed thanks to the deep (28 AB magnitude in Near-infrared for a signal/noise=10 in 10$^4$ s\footnote{https://jwst.stsci.edu/science-planning/proposal-planning-toolbox/sensitivity-overview}) for spectroscopy that JWST will perform in optical and near-infrared. Template-fitting methods will then have to incorporate the new knowledge in form of new templates and new physics.

\subsection{Building blocks to unveil new cosmology}

The next decade will see many exciting imaging surveys dedicated to cosmology. The goal of these surveys is to constrain the nature of the dark energy through the combination of various probes. Several probes rely on photo-z, such as galaxy cluster counts, weak lensing tomography, and Baryon Acoustic Oscillations (BAO). These probes require large statistical samples over large areas to extract their cosmological signals, which has triggered this period of gigantism for the next generation of cosmological surveys. While the ongoing DES survey will cover 5000 deg$^2$ in 5 bands and gathering information for 300 millions of sources, the next generation of surveys will increase the number of galaxies by one order of magnitude. For instance, LSST \citep[][]{Ivezic:2008aa} will begin operations in 2021 and cover 18000 deg$^2$ of the sky,  gathering information on 4 billions sources down to $r_{AB}=27.5$ after 10 years of operation. The Euclid mission \citep[][]{Laureijs:2011aa} will be launched in 2021 and remain in operation for 6 years and use its VIS camera to measure the galaxy shapes of 1.5 billion of galaxies over 15000 deg$^2$. 
 
Computation of photo-z and storage of the PDF will be as  challenging as to maintain the required photometry quality across such a large area. For instance, one requirement of the LSST survey is to get band-to-band calibration errors not larger than 0.005 mag and no more than 0.01 mag variation across the sky. Being successful in maintaining such relative calibration is necessary to insure homogeneous performances.

As describe in section \ref{sec:validation}, the multi-lambda coverage is also crucial to define the redshift range of interest. In the new generation of imaging surveys, unless SPHEREx \citep{Dore:2016aa} with its 96 medium band NIR photometry is approved, Euclid will be the only survey to map the Universe in NIR using three filters between 9200\AA{}-20000\AA{}. The goal of these filters is to insure precise photometric redshifts at $z>1.3$ which is not possible without NIR (see panel a) of Fig.~\ref{fig:degeneracy}). However, Euclid will need to be complemented with ground-based data in optical wavelengths.

While most of the cosmological imaging surveys are performed with broad bands (HSC, DES, LSST, Euclid), two surveys are performed using medium bands: the PAU and the J-PAS surveys. For instance, J-PAS \citep[][]{Benitez:2014aa} will cover 8500 deg$^2$ with 54 narrow band filters. They should observe 300 million galaxies with a photo-z precision of 0.3\%. With this precision one of the main objectives of this survey will be to the measure the BAO.

Finally, the variable Universe will become more easily accessible with LSST due to the 1000 repeated observations of each location of the sky over 10 years, starting in 2022, in up to six bands. In particular it will include the $u$ band, assuring the capability of breaking the degeneracy between low and high redshift solutions. More in general, LSST will be extremely powerful to break the identify stars, AGN and exotic sources. 

\subsection{Evolution in the photo-z technique and synergy with spectroscopic surveys}

The evolution of the photo-z methods will depend on the spectroscopic data that will become available. For the next generation of cosmological surveys, the imaging and spectroscopic surveys are often conceived in parallel. For instance, the HSC imaging survey will be complemented with spectroscopy using the multi-object Prime Focus Spectrograph (PFS) instrument covering a wavelength range 3800\AA{}-13000\AA{}, and gathering redshift for millions of galaxies at $0.8<z<6$ \citep[][]{Tamura:2016aa}. Similarly, the Euclid survey is complemented by the spectroscopy performed by the NISP instrument, able to perform NIR (11000\AA-20000\AA) slit-less spectroscopy to detect the $H_{\alpha}$ emission for more than 50 millions of galaxies at $0.7<z<2$. With such a bright future for spectroscopic surveys, the ML techniques will  become competitive in the high redshift Universe. It must also be stressed that because of the incumbent massive photometric surveys designed as cosmological probes, the algorithms and techniques for computing reliable photo-z are in continuous development. The most recent methods are hybrids that combines the best of the ML and SED fitting techniques \citep[e.g.,][]{Vanzella:2004aa, Carrasco-kind:2014bb, Beck:2016aa, Leistedt:2017aa, Speagle:2017aa}.

\begin{figure*}
\centering
\includegraphics[scale=0.5,clip=true,trim=10 10 10 10]{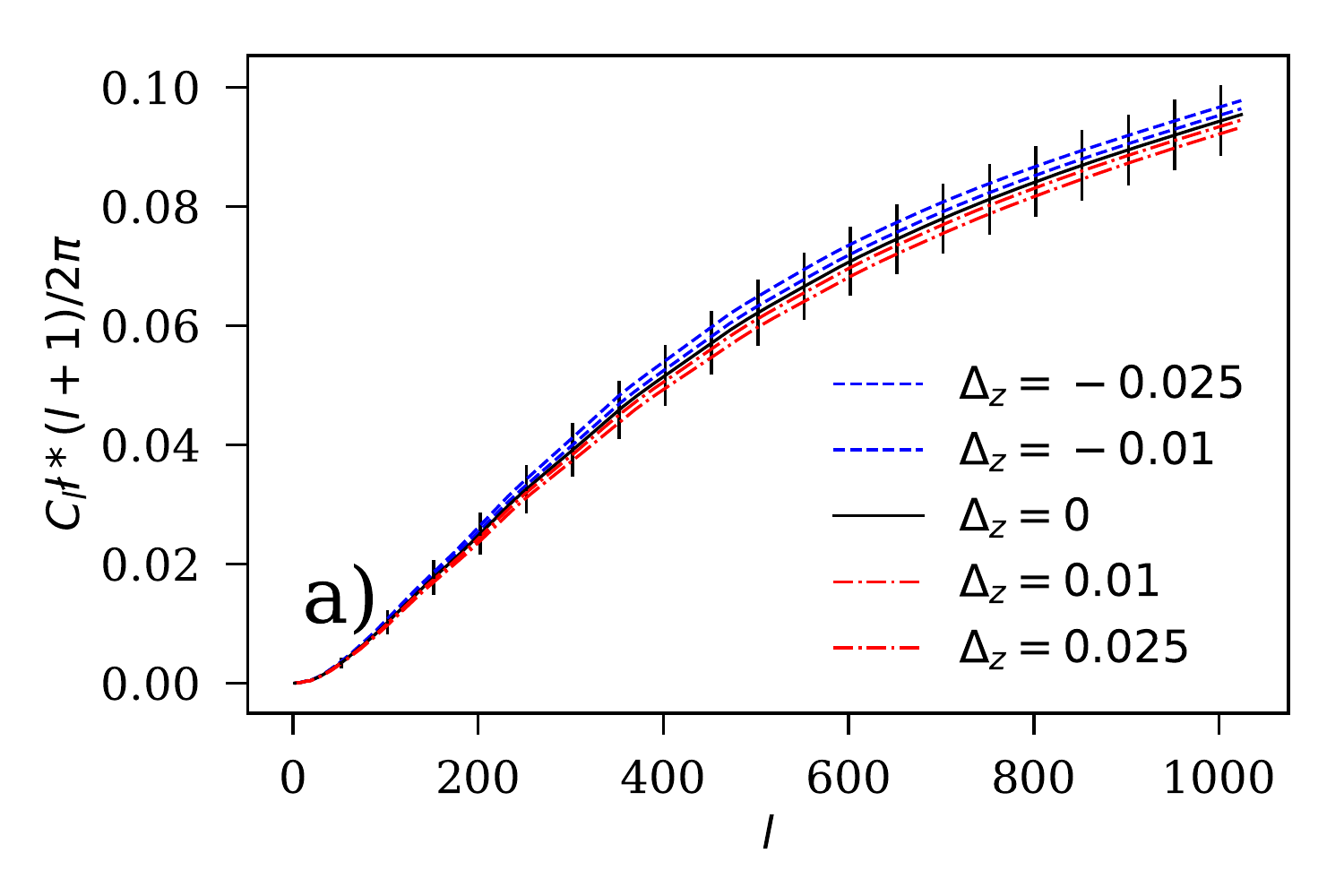}
\includegraphics[scale=0.5,clip=true,trim=10 10 10 10]{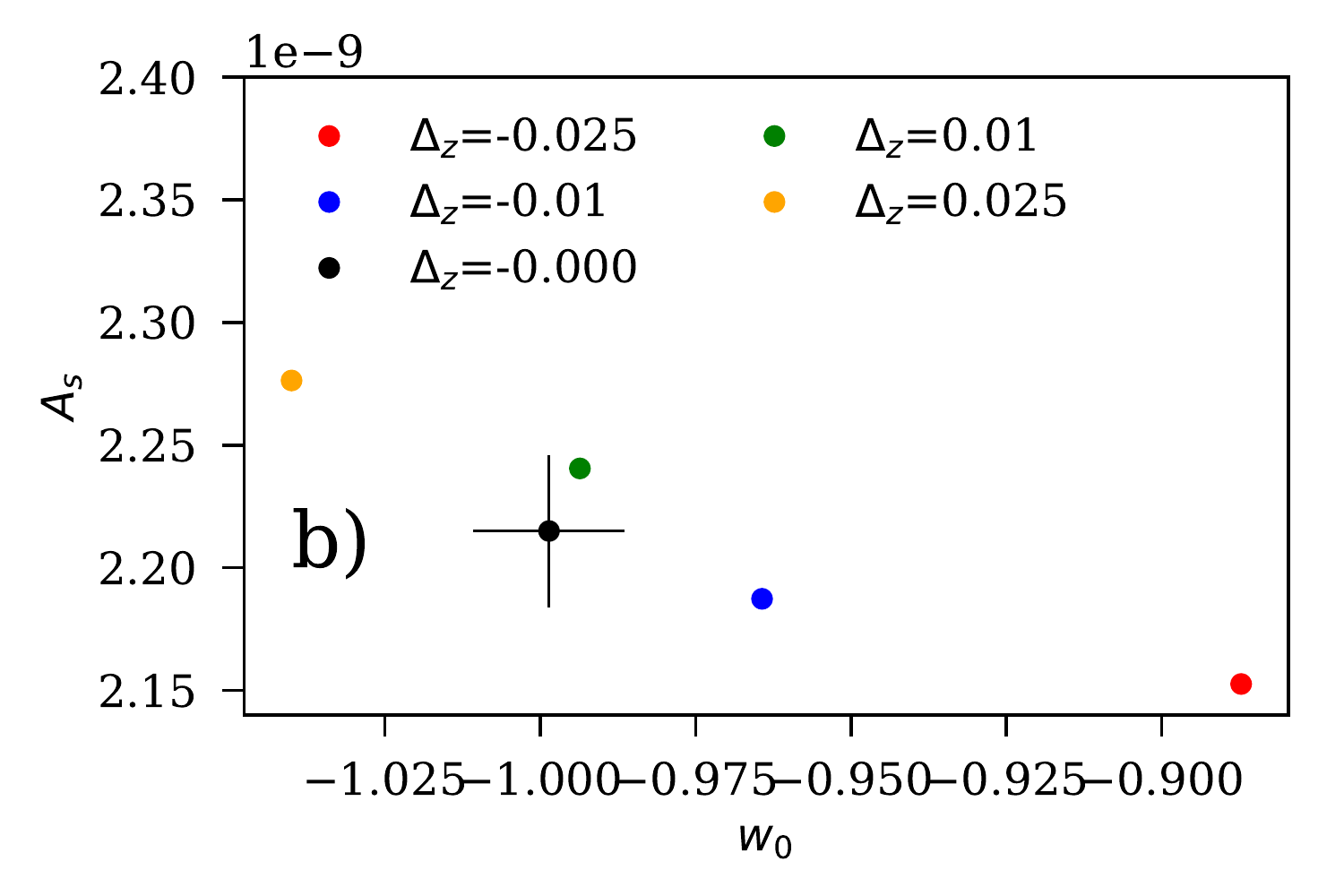}
\caption{ {\bf Cosmological parameter biases incurred by photo-z inaccuracy. a):} the predicted auto-correlation function ($C_l$) is shown for a simulated Euclid-like redshift distribution (black line) and for a redshift distribution modified by a translation of up to $\pm0.025$ (red and blue lines). {\bf b)}: The subsequent biases induced in the cosmological parameters, $w_0$ - the equation of state of dark energy, and $A_s$ - the amplitude of the fluctuations on the surface of last scattering, from the redshift translations.}
\label{fig:science_application}
\end{figure*}

Specific needs will also trigger new techniques. Scientific application based on the weak lensing tomography require the characterisation of the true redshift distribution with an extreme precision in order to map the result of a weak lensing shear analysis onto cosmological parameters. In a very simplified setting, we show in Fig.\ref{fig:science_application} how a biased photo-z distribution expected for one tomographic redshift bin for a Euclid-like survey, will impact the predicted auto-correlation function, and therefore will lead to incorrect best fitting cosmological parameters. The actual effect of such an offset depends strongly on the redshift of the galaxy sample, and the full shape of the redshift distribution function. In a weak lensing setting, we direct the reader to \cite[Fig.1.3 in ][]{Newman:2015aa, Amendola:2013aa}  showing how a bias on the mean redshift translates into an error on the time evolution component of the dark energy equation of state in the LSST case. With current methods, such low bias can not be achieved using photo-z alone. The development of new techniques, able to combine the photo-z with the sources position (see \S.~\ref{subsec:spatial}) have the potential to revolutionise the field.

%----------------------------
%----------------------------

\acknowledgements{The Authors are grateful to the referees and the editor who contribute to the review with their comments. The authors are grateful to  Li-Ting Hsu, Jonas Chavez-Montego, Pablo Gonzalez, Hendrik Hildebrandt, Massimo Brescia and Stefano Cavuoti for providing the CANDELS/CDFS, ALHAMBRA, SHARDS, KiDS\_BPT, KiDS\_MLPQN data that were used in Fig.~\ref{fig:accuracy}.}
%-----------------------

\end{document}